\documentclass[%
 aip,
 sd,%
 amsmath,amssymb,
 preprint,%
]{revtex4-1}

\usepackage{graphicx}
\usepackage{dcolumn}
\usepackage{bm}
\usepackage[colorlinks=true,linkcolor=blue,citecolor=blue]{hyper ref}
\usepackage{natbib}

\begin{document}

\title {Dynamic Actuation of Single-Crystal Diamond Nanobeams}

\author{Young-Ik Sohn}
\author{Michael J. Burek}
\affiliation{Paul A. Johnson School of Engineering and Applied Sciences, Harvard University, 29 Oxford Street, Cambridge, Massachusetts 02138, United States}
\author{Vural Kara}
\affiliation{Department of Mechanical Engineering, Division of Materials Science and Engineering, and the Photonics Center, Boston University, Boston, Massachusetts 02215, USA}
\author{Ryan Kearns}
\affiliation{Paul A. Johnson School of Engineering and Applied Sciences, Harvard University, 29 Oxford Street, Cambridge, Massachusetts 02138, United States}
\affiliation{Department of Chemical Engineering, University of Waterloo, 200 University Avenue West, Waterloo, Ontario, Canada,  N2L 3G1}
\author{Marko Lon\v{c}ar}
\email{loncar@seas.harvard.edu.}
\affiliation{Paul A. Johnson School of Engineering and Applied Sciences, Harvard University, 29 Oxford Street, Cambridge, Massachusetts 02138, United States}


\begin{abstract}
We show the dielectrophoretic actuation of single-crystal diamond nanomechanical devices. Gradient radio-frequency electromagnetic forces are used to achieve actuation of both cantilever and doubly clamped beam structures, with operation frequencies ranging from a few MHz to $\sim$50MHz. Frequency tuning and parametric actuation are also studied.
\end{abstract}

\maketitle

Owing to its large Young's modulus, excellent thermal properties, and low thermoelastic dissipation, \textit{single-crystal} diamond (SCD) is a promising candidate for realization of high frequency (\textit{f}) and high quality factor (\textit{Q}) mechanical resonators. Indeed, significant advances in diamond fabrication have made it possible to achieve mechanical $Q$-factors exceeding 1 million at room temperature for micron scale SCD mechanical resonators.\cite{Tao:2014jh} Such devices are of interest for realization of stable, high \textit{f} $\cdot$ \textit{Q} product, radio frequency (RF) oscillators for inertial sensing applications.\cite{Kusterer:2009uw} SCD is also a promising platform for applications in quantum information science and technology due to the color centers which can be embedded inside.\cite{Rabl:2010kk} In particular, the negatively charged nitrogen vacancy (NV) color centers can be used as qubits with optical readout due to their long coherence times (milliseconds) even at room temperature.\cite{Balasubramanian:2009fu} For example, coupling between an NV center and a mechanical resonator may enable high fidelity control of NV center spin state via rapid adiabatic passage,\cite{MacQuarrie:2013cp,MacQuarrie:2015bx} and potentially the remote coupling of distant NV centers via mechanics.\cite{Rabl:2010kk} Finally, mechanical resonators may enable coherent coupling between systems with degrees of freedom possessing dramatically different properties and energy scales. 

Here, we demonstrate nanoscale resonators with high \textit{f} $\cdot$ \textit{Q} product in SCD. To drive the resonators we use  dielectrophoretic actuation,\cite{Unterreithmeier:2009gh} which allows us to realize nanoelectromechanical systems (NEMS) at a frequency range of 1-50 MHz with flexural mechanical modes.
\begin{figure}[h]
        \centering
        \includegraphics[width=\columnwidth]{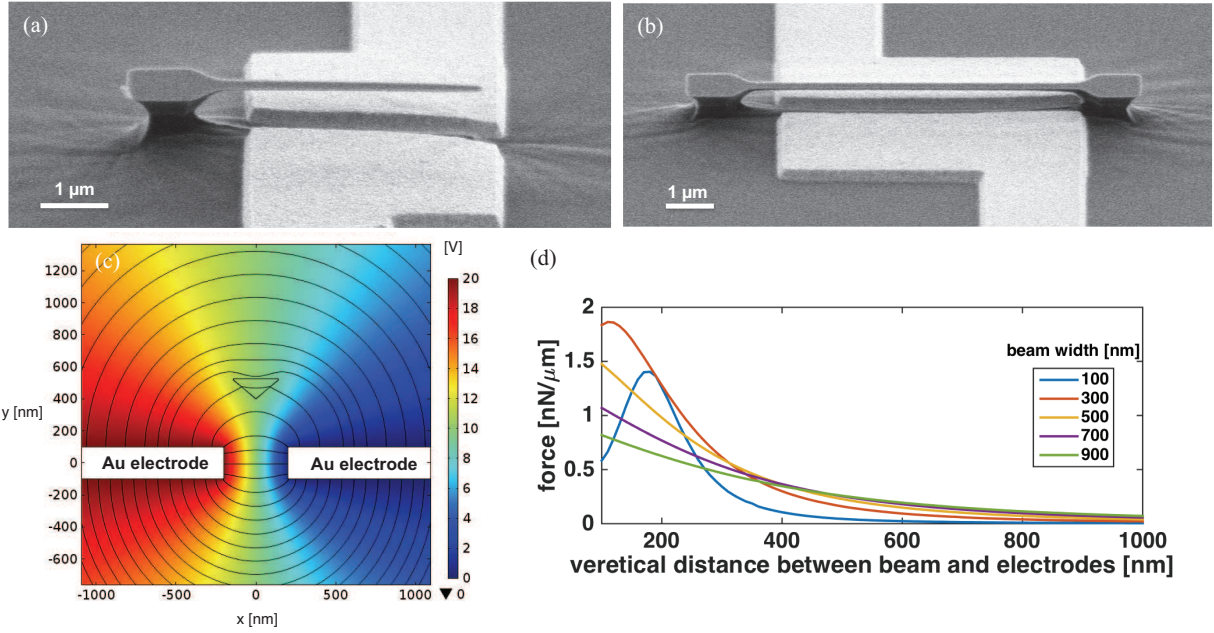}
        \caption{SEM images of (a) 4 $\mu$m long cantilever and (b) 7 $\mu$m doubly clamped beam. (c) Finite element method (FEM) simulations are used to calculate the force applied to suspended nanobeams with a given geometry and electrostatic environment. The color map indicates potential with respect to the right-hand Au electrode and the streamlines show the corresponding electric field. (d) Vertical force per unit length applied to such beams in the case of 20V of DC voltage is plotted as a function of beam width and distance above the electrode. Separation between electrodes is the sum of beam width and 50 nm margin on either side. Beam height is the distance between top surface of the beam and the electrode center in vertical axis.}
        \label{FigureOne}
\end{figure}
Dielectrophoresis has been used in the past to achieve mechanical resonance tuning,\cite{Rieger:2012gu} coherent control of classical mechanical resonators,\cite{Faust:2013gs} cavity electromechanics,\cite{Faust:2012kj} and nonlinear mechanics.\cite{Unterreithmeier:2010eo} In our approach, on-chip metal electrodes are fabricated on either side of SCD nanobeam cantilevers (Fig.  \ref{FigureOne}(a)) and doubly clamped nanobeams (Fig. \ref{FigureOne}(b)). Fringing electromagnetic fields of an RF drive the diamond devices (Fig. \ref{FigureOne}(c)), with optimal actuation occurring when the RF frequency is resonant with the mechanical mode. Our numerical modeling indicates that it is crucial that the vertical distance between the metal electrodes and diamond nanobeam is small in order to achieve efficient actuation (Fig. \ref{FigureOne}(d)). 

Other actuation schemes for nanomechanical resonators have been demonstrated previously, including electrostatic and piezo-electric actuation approaches.\cite{Ekinci:2005dg} These, however, require deposition of a conductive thin film or electronic doping on the moving part of nanomechanical structure, because undoped diamond is neither conductive nor piezoelectric. These can reduce mechanical $Q$-factors\cite{Imboden:2014vf}  and negatively impact spin and optical degrees of freedom of color centers embedded inside diamond. The latter are known to be sensitive to the fabrication imperfections and surface terminations.\cite{Chu:2014vm} 
Forces resulting from gradient electromagnetic fields, on the other hand, do not require any modifications to the diamond mechanical resonator. Therefore, the dielectrophoresis scheme does not add additional mechanical loss channels. 
One caveat is that careful design of device geometry is required for the dielectrophoretic actuation, because its force at a given voltage is much weaker than other methods.

\begin{figure}[ht]
        \centering
        \includegraphics[width=\columnwidth]{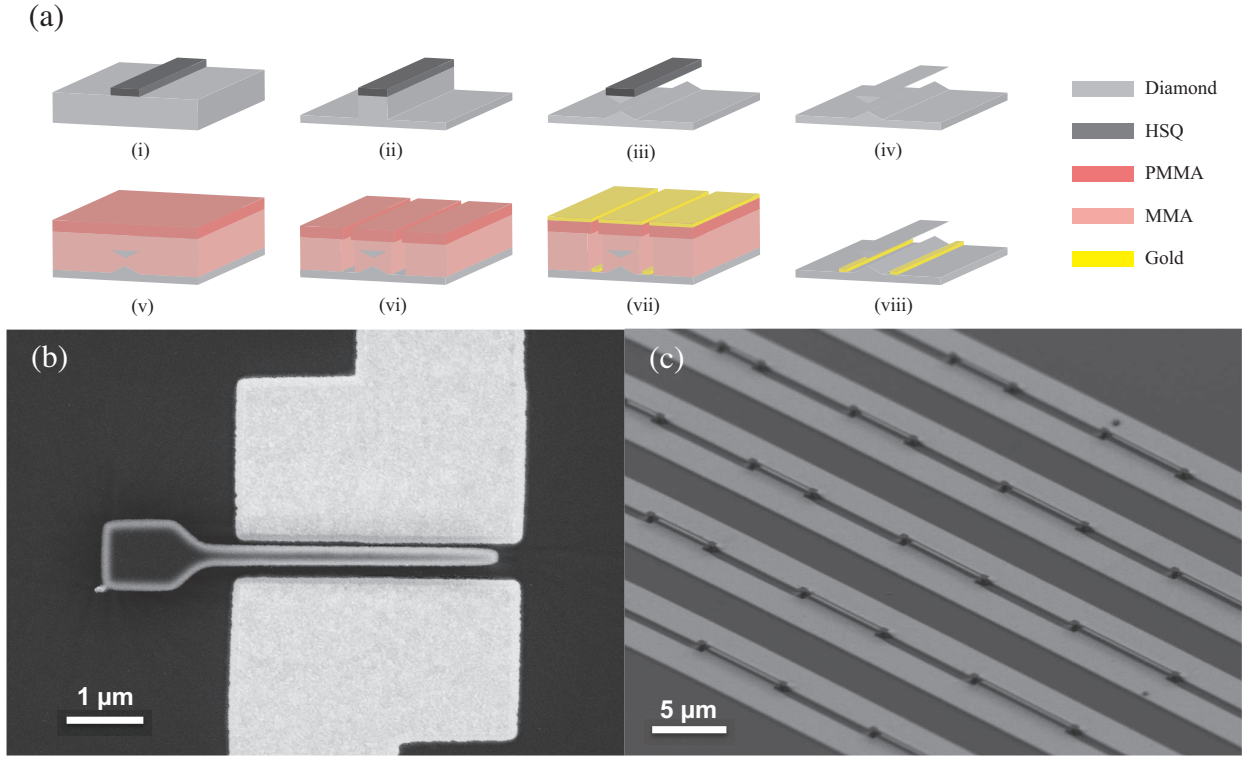}
        \caption{(a) Schematic illustration of angled-etching nanofabrication approach used in this work:(i) Electron beam lithography mask is deposited, (ii) top-down reactive ion etching of diamond is performed, followed by the (iii) angled-etching step and (iv) mask removal. (v) New electron beam resist is spin coated, and (vi) electron beam lithography followed by (vii) metal evaporation and (viii) lift-off are used to define electrodes. (b) High magnification SEM image of 4 $\mu$m cantilever shows that good alignment can be achieved. (c) SEM image of device array sharing electrodes.}
        \label{FigureTwo}
\end{figure}

The fabrication scheme for realizing diamond NEMS is shown in Fig. \ref{FigureTwo}(a). Diamond nanocantilevers and doubly clamped nanobeams are first fabricated using our recently developed angled-etching technique, described in detail elsewhere.\cite{Burek:2012bk} Briefly, angled-etching employs anisotropic oxygen plasma etching at an oblique angle to the substrate surface, yielding suspended triangular cross-section nanobeams directly from single-crystal bulk diamond substrates. To ensure efficient actuation by dielectrophoresis, the diamond nanobeam width and distance between the substrate and the bottom apex of the triangular nanobeam cross-section must be carefully chosen (Fig. \ref{FigureOne}(c) and (d)). Once free-standing diamond nanobeams are fabricated, metal electrodes are patterned on the diamond substrate via lift-off process. First, the diamond substrate is spin coated with a polymethylmethacrylate-copolymer (MMA/PMMA) bilayer resist, where the MMA copolymer thickness is chosen to be slightly thicker than the distance between the nanobeam top surface and the substrate. After resist coating, exposure and alignment are done with electron beam lithography. After developing the resist, an adhesion layer of 50 nm titanium and a 200 nm thick gold layer are evaporated on the surface by electron beam evaporation. Lift-off in Remover PG completes electrode patterning. Fig. \ref{FigureTwo}(b) is a top-down SEM image of a diamond nanobeam cantilever with gold electrodes fabricated on either side. We observe very good alignment of the electrodes to the diamond nanobeam, with alignment errors on the order of tens of nanometers. In fact, the slight misalignment enables the actuation of diamond nanobeam in-plane motion.\cite{Rieger:2012gu} Fig. \ref{FigureTwo}(c) shows an array of fabricated diamond doubly clamped nanobeam mechanical resonators that share driving electrodes. This configuration allows us to characterize in parallel a large number of resonators having slightly different geometry and hence different mechanical resonance frequencies.\footnote{This array of devices made on a separate chip with different electrode configurations from other devices presented in this work.} Our diamond nanomechanical resonators had a width range between 200 nm and 300 nm and lengths between 1 $\mu$m and 20 $\mu$m, corresponding to fundamental flexural resonance frequencies ranging from a few MHz to hundreds of MHz. We note that due to the nature of our angled-etching fabrication technique, the width and thickness of the nanobeam triangular cross-section are correlated. \cite{Burek:2013bq}

\begin{figure}[ht]
        \centering
        \includegraphics[width=\columnwidth]{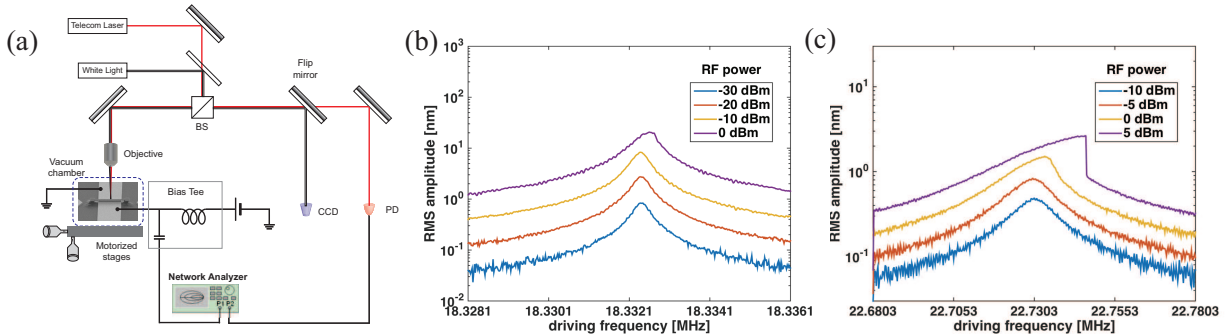}
        \caption{(a) Optical characterization setup. Fundamental out-of-plane resonant response of devices shown in Fig. \ref{FigureOne}(a) and (b) are given in (b) and (c), respectively. Lorentzian frequency responses are shown at low driving power, and both beams start to enter nonlinear regime at higher driving power.}
        \label{FigureThree}
\end{figure}

All experiments were performed at room temperature, with our wire bonded diamond substrate in a vacuum chamber, held at a pressure below $10^{-4}$ Torr. Fig. \ref{FigureThree}(a) shows a schematic of the optical interferometry characterization setup\cite{Ekinci:2005dg} used to read out the nanomechanical motion.\footnote{One particular experiment (Fig. \ref{FigureThree}(b)) was performed with the separate setup, which is a path-stabilized Michelson interferometer. The reason is to have the same sensitivity with less incident laser power, which makes the measurement more stable. In this setup, He--Ne laser with wavelength ($\lambda \approx$ 632 nm) was used.} Sending RF signals for actuation and read-out at corresponding frequencies was done in a transmission measurement with a Vector Network Analyzer (VNA). The VNA was replaced with a real-time spectrum analyzer for those measurements (e.g. measuring thermal fluctuation of the nanobeam) which did not involve actuation, and the parametric actuation measurement that we discuss later. A bias-tee was also included to combine a DC bias with the RF drive signal to ensure proper actuation, as the dielectrophoresis actuation force is proportional to the square of applied voltage, $F  \propto \left(V_{DC}+V_{RF} \cos{\omega t}\right)^2$.\cite{Unterreithmeier:2009gh}

For the most of fabricated diamond nanobeams, both the fundamental out-of-plane mechanical flexural modes were characterized. Resonant responses of the fundamental out-of-plane motion of devices shown in Fig. \ref{FigureOne}(a) and (b) are plotted in Fig. \ref{FigureThree}(b) and (c), respectively. Curves are the raw data, with both figures showing the expected resonant responses at low driving power as well as nonlinear response at higher driving power. 10V of DC voltage was applied for the both measurements. The resonance frequency of out-of-plane mode that we could measure on 4 $\mu$m long cantilever was $\sim$18.3 MHz with the mechanical quality factor of $4.4\times10^4$. In the case of 7 $\mu$m long doubly clamped nanobeam, measured resonant frequency was $\sim$22.7 MHz with the mechanical quality factor of $2.0\times10^3$. For both devices, root mean square (RMS) amplitude of motion was thermomechanically calibrated by measuring thermal fluctuations.\cite{Hauer:2013db} To do so, the value of effective mass is estimated from nanobeam's geometry and analytic theory presented elsewhere.\cite{Hauer:2013db} The width and length of nanobeams are measured via SEM, and the thickness is estimated from the ratio of out-of-plane and in-plane resonance frequencies of the cantilever.\cite{Burek:2013bq}

Highest resonant frequency measured with driven motions is as large as $\sim$50 MHz, although its thermal fluctuation was not detectable. To the best of our knowledge, this is the highest actuation frequency of flexural mechanical vibration achieved by dielectrophoretic actuation to date. Unfortunately, in our current experiments, we were not able to measure devices with resonances $>$50 MHz, due to the limited sensitivity of our measurements ($\sim$ 0.5 pm/$\sqrt{\text{Hz}}$). In our current characterization setup, the noise floor of our detection was affected by three different instruments: shot noise from laser source, dark current and thermal noise from the photodetector and thermal noise from the receiver. Depending on the settings of instruments, any of these three could be the limiting factor for the detection noise floor. \footnote{When maximum telecom laser power of 20 mW was used, amount of power reaches the sample is $<$ 5mW. In such a case, the setup had a sensitivity of $\sim$ 0.5 pm/$\sqrt{\text{Hz}}$. For the path-stabilized Michelson interferometer, when $<$ 400 $\mu$W reached the sample, it gave a similar sensitivity of $\sim$ 0.5 pm/$\sqrt{\text{Hz}}$. Signal-to-noise ratio can be always improved with the higher laser power, however, it accompanies the heating of the device and unstable measurement.}

In many MEMS / NEMS applications, a high \textit{f} $\cdot$ \textit{Q} product is the key figure of merit. For example, in the case of mass sensing based on mechanical resonator, sensitivity scales with the square of its frequency and the quality factor determines the minimum detectable frequency shift.\cite{Rinaldi:fa}
State-of-the-art flexural NEMS device can reach \textit{f} $\cdot$ \textit{Q} product of $6.8 \times 10^{12}$ Hz.\cite{Verbridge:2008ja} In our devices, the maximum \textit{f} $\cdot$ \textit{Q} product that we measured was $8.1 \times 10^{11}$ Hz in the case of a 260 nm wide and 4 um long diamond nanobeam cantilever (Fig. \ref{FigureThree}(b)). 

For applications in quantum information science, coupling of NV center with mechanical resonator has been studied recently with various platforms.\cite{MacQuarrie:2013cp,MacQuarrie:2015bx,Ovartchaiyapong:2014gv,Teissier:2014gt} Assuming our device in Fig \ref{FigureThree}(b) is used for such experiment, we estimate the relevant physical quantities below.
When RF power of $-10$ dBm is applied on its resonance, assuming that NV is implanted near the clamp at 10 nm depth, applied strain at the site of it is estimated to be $7.4\times10^{-5}$ from FEM modeling. Estimated strain is large enough to induce significant coupling of NV ground-state spin with mechanical vibrations. For example, if z-axis of NV is perpendicular to the length direction of the cantilever, estimated strain corresponds to the coupling of 1.6 MHz.\cite{Ovartchaiyapong:2014gv} 

\begin{figure}[ht]
        \centering
        \includegraphics[width=\columnwidth]{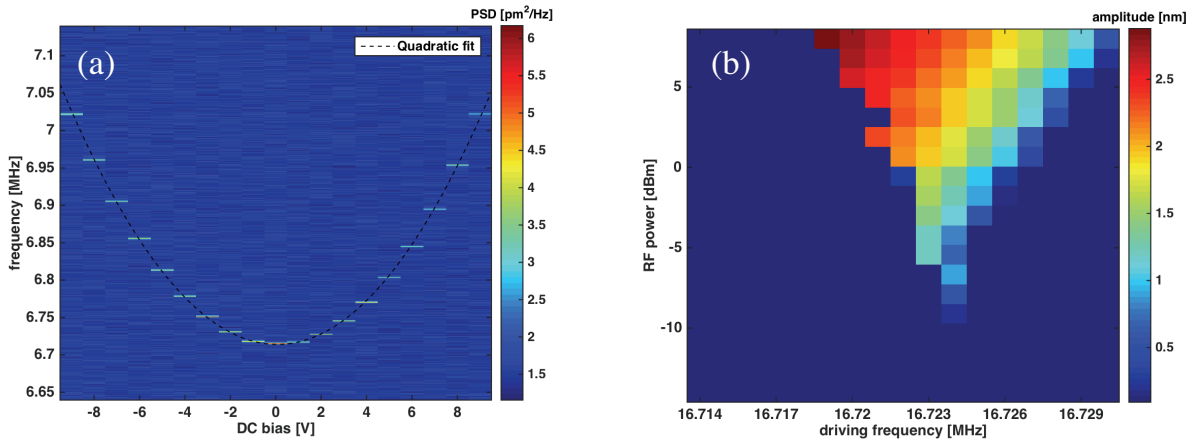}
        \caption{(a) Tuning of mechanical resonance of doubly clamped beam using DC bias. With applying $\pm 9$V, frequency tuning range that can be achieved is approximately 260 linewidths. (b) Typical tongue shape of parametric instability was observed.}
        \label{FigureFour}
\end{figure}

In addition to basic actuation capability, the dielectrophoretic actuation scheme can be used to tune the mechanical resonance frequency.\cite{Unterreithmeier:2009gh} This is because the actuation force has dependence on the displacement of the diamond nanobeam. Since the force has quadratic dependence on applied voltage, the amount of shift in the resonance frequency has quadratic dependence as well. Fig. \ref{FigureFour} (a) shows power spectral density (PSD) of the thermal fluctuations of a doubly clamped nanobeam (250 nm wide, 100 nm thick, 19 $\mu$m long), as the applied DC bias was changed from $-9$ V to $+9$ V. Bright spots observed in each data column correspond to the resonance frequencies. The solid black line is a quadratic fit for applied DC bias and shows an excellent match with the data. In the given range of applied DC bias, the mechanical resonance could be tuned over roughly 260 full widths at half maxima of the resonance peak. We observed a blue shift of the diamond nanobeam resonance when DC voltage is applied which differs from observed red shift in the similar work.\cite{Unterreithmeier:2009gh} Upon further inspection, it was observed that our nanobeams are buckled down due to  considerable amount of  residual compressive stress (due to the diamond growth process).\cite{Friel:2009ht} See supplemental material at [URL will be inserted by AIP] for the SEM image of buckled doubly clamped beam used in the measurement. Therefore, the central part of nanobeams are positioned quite close to the top surface of the electrodes than design value, in which case the blue shift in resonance is expected.\cite{Rieger:2012gu}
 
Since the resonance frequency is easily parametrically tuned much more than a linewidth, parametric excitation is also expected. When the spring constant of nanobeam is a function of the displacement, its motion can be modeled by Mathieu's equation as shown below:
\begin{gather}
	\left[ \frac{d^2}{dt^2}+\frac{\Omega_0}{Q}\frac{d}{dt}+\Omega_0^2\left(1+\alpha-2\Gamma \sin{2\Omega_0 t}\right) \right]x(t)=0
	\label{Mathieu's equation}
\end{gather}
where $x(t)$, $\Omega_0$, $Q$, $F(t)$ and $m$ are the beam displacement, the mechanical resonance frequency, mechanical $Q$-factor, external driving force and effective mass of the resonator, respectively. $\alpha$ is the detuning from parametric excitation and  $\Gamma$ is proportional to the parametric excitation amplitude. The criteria for the onset of parametric instability is $\Omega_0/Q=\Gamma$.\cite{Mahboob:2008dk} Mathieu's equation can be analytically solved and the solution predicts its stability on a phase plane, axes of which are detuning and driving amplitude. Here, we show an ``instability tongue''\cite{Nayfeh:2008wl}  when a doubly clamped diamond nanobeam is parametrically excited. In Fig. \ref{FigureFour}(b) the measured instability tongue is shown when the nanobeam (250 nm wide, 100 nm thick, 16 $\mu$m long) was excited around twice its natural frequency of $\sim$8.36 MHz, with 10 V of DC voltage  applied together. In this experiment, excitation was applied by an RF signal generator and the response was measured with spectrum analyzer, with the amplitude of the driven motion thermomechanically calibrated.\cite{Hauer:2013db} Parametric excitation is particularly interesting for NEMS devices since it can circumvent electric cross talk, which can be detrimental for nanoscale systems,\cite{Feng:2008ca} and can be used to realize a NEMS oscillator\cite{Villanueva:2011bw} and mechanical memory element.\cite{Mahboob:2008dk}

In summary, we have realized a resonant actuator based on dielectrophoresis for SCD nanomechanical resonators. Actuation of both cantilever and doubly clamped diamond nanobeams was achieved for both the fundamental out-of-plane vibrations. Our driving frequency range spanned from a few MHz to nearly 50 MHz, though higher frequency actuation is expected to be measured by a displacement read-out scheme with better sensitivity. Additional functionalities of the system are frequency tuning with DC bias and parametric excitation. The SCD actuation scheme we developed here is expected to be an excellent platform for coupling NV energy levels to mechanical degree of freedom. Additionally, control over diamond nanobeam mechanical motion by dielectrophoresis forces may be applied in the resonance tuning and modulation of recently demonstrated diamond optical cavities, in a manner similar to what has previously been demonstrated with silicon nanophotonic devices.\cite{Frank:2010tv,Deotare:2012cp,Burek:2014bj}

The authors would like to thank S. Meesala for helpful discussions. This work was supported by the STC Center for Integrated Quantum Materials, NSF Grant No. DMR-1231319, the Defense Advanced Research Projects Agency (QuASAR program), and AFOSR MURI (grant FA9550-12-1-0025). M.J. Burek is supported in part by the Natural Science and Engineering Council (NSERC) of Canada. This work was performed in part at the Center for Nanoscale Systems (CNS), a member of the National Nanotechnology Infrastructure Network (NNIN), which is supported by the National Science Foundation under NSF award no. ECS-0335765. CNS is part of Harvard University.

\bibliography{myrefs}

\end{document}